\documentclass[10pt]{article}
\usepackage{color}
\usepackage{graphicx}
\usepackage{url}
\usepackage[final]{trackchanges}
\renewcommand{\initialsOne}{GH}
\renewcommand{\initialsTwo}{TZ}
\renewcommand{\initialsThree}{GW}
\usepackage{listings}
\newcommand{\comment}[1]{}

\newcommand{\Section}{\section}
\newcommand{\SubSection}{\subsection}
\begin{document}
\title{Data Access Optimizations for Highly Threaded Multi-Core CPUs
with Multiple Memory Controllers}
\author{Georg Hager\thanks{\protect\url{georg.hager@rrze.uni-erlangen.de}}\qquad Thomas Zeiser\thanks{\protect\url{thomas.zeiser@rrze.uni-erlangen.de}}\qquad Gerhard Wellein\thanks{\protect\url{gerhard.wellein@rrze.uni-erlangen.de}}\\
Regionales Rechenzentrum Erlangen\\ 91058 Erlangen, Germany}
\date{January 28, 2008}
\maketitle
\begin{abstract}
Processor and system architectures that feature multiple
memory controllers are prone to show bottlenecks and erratic
performance numbers on codes with regular access
patterns. Although such effects are well known in the form of
cache thrashing and aliasing conflicts, they become more
severe when memory access is involved. Using the new Sun
UltraSPARC T2 processor as a prototypical multi-core design,
we analyze performance patterns in low-level and application
benchmarks and show ways to circumvent bottlenecks by careful
data layout and padding.

\end{abstract}


\Section{The Sun UltraSPARC T2 processor}

\begin{figure}[tbp]
\includegraphics[width=\columnwidth]{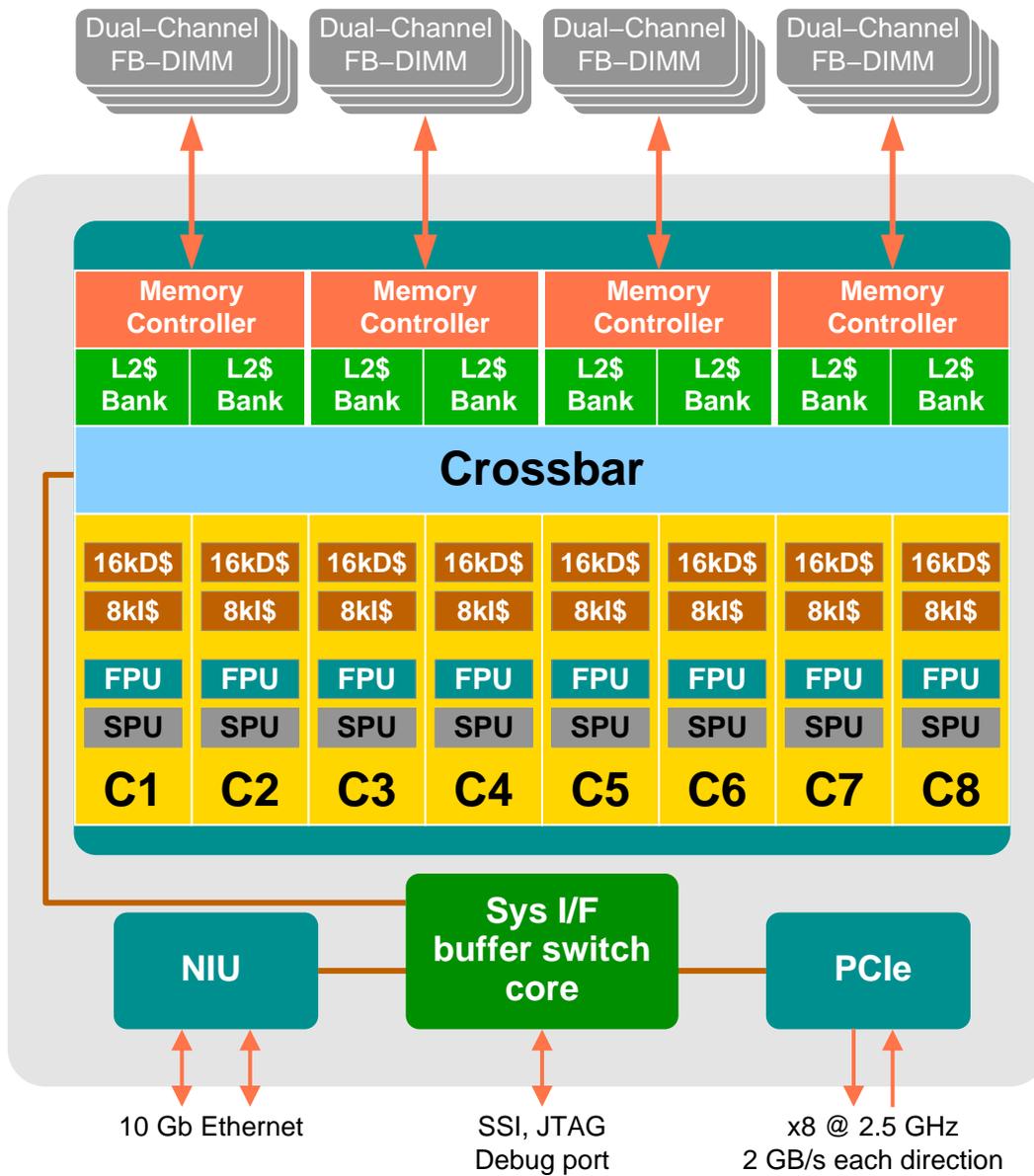}%
\caption{\label{fig:niagara-arch}Block diagram of the Sun UltraSPARC T2 
processor (see text for details). Picture by courtesy of Sun Microsystems.}
\end{figure}

Trading high single core performance for a highly parallel single
chip architecture is the basic idea of T2 as can be seen in
Fig.~\ref{fig:niagara-arch}: Eight simple in-order SPARC cores
(running at 1.2 or 1.4\,GHz) are connected to a shared, banked L2
cache and four independently operating dual channel FB-DIMM memory
controllers through a non-blocking switch, thereby providing UMA
access characteristics with scalable bandwidth. Such features were
previously only available in shared-memory vector computers like the
NEC SX series. To overcome the restrictions of in-order architectures
and long memory latencies, each core is able to support up to eight
threads, i.e.\ there are register sets, instruction pointers etc.\ to
accommodate eight different machine states. There are two integer, two
memory and one floating point pipeline per core. Although all eight
threads can be interleaved across the floating point and memory pipes,
each integer pipe is hardwired to a group of four threads. The CPU can
switch between the threads in a group on a cycle-by-cycle basis, but
only one thread per group is simultaneously active at any time. If a
thread has to wait for resources like, e.g., memory references, it
will be put in an inactive state until the resources become available
which allows for effective latency hiding \cite{t2tech} but restricts
each thread to a single outstanding cache miss. Running
more than a single thread per core is therefore mandatory for most
applications, and thread placement (``pinning'') must be implemented.
This can be done with the standard Solaris \verb.processor_bind(). system call
or, more conveniently but only available for OpenMP, using the
\verb.SUNW_MP_PROCBIND. environment variable.

Each memory controller is associated with two L2 banks. A very simple
scheme is employed to map addresses to controllers and banks: Bits 8
and 7 of the physical memory address select the memory controller to
use, while bit 6 determines the L2 bank
\cite{t2tech,hether}\@. Consecutive 64-byte cache lines are thus served
in turn by consecutive cache banks and memory controllers.  Due to the
fact that typical page sizes are at least 4\,kB the distinction
between physical and virtual addresses is of no importance here.

The aggregated nominal main memory bandwidth of 42\,GB/s (read) and
21\,GB/s (write) for a single socket is far ahead of most other
general purpose CPUs and topped only by the NEC SX-8 vector
series. Since there is only a single floating point unit (performing
MULT or ADD operations) per core, the system balance of approximately
4 bytes/flop (assuming read) is the same as for the NEC SX-8 vector
processor. \change[GH]{In practice}{In our experience, as shown in 
Sect.~\ref{sec:stream},} only about one third of the theoretical bandwidth
can actually be measured. 

Beyond the requirements of the tests presented here one should be
aware that the T2 chip also comprises on-chip PCIe-x8 and 10\,Gb
Ethernet connections as well as a cryptographic coprocessor. These
features are reminiscent of the actual concept of the chip: It is
geared towards commercial, database and typical server workloads.
Consequently, one should not expect future versions to improve on
HPC-relevant weaknesses of its design.

\Section{Benchmarks and optimizations}

This section describes the benchmarks that were used to pinpoint
aliasing effects, performance results and optimization techniques.
All measurements were performed on a Sun SPARC Enterprise T5120 system
running at 1.2\,GHz\@.

\SubSection{McCalpin STREAM}\label{sec:stream}

The STREAM benchmark \cite{stream} is a widely used code to assess
the memory bandwidth capabilities of a single processor or shared memory
computer system. It performs the OpenMP-parallel operations
\begin{itemize}
\item copy: \verb.C(:)=A(:).
\item scale: \verb.B(:)=s*C(:).
\item add: \verb. C(:)=A(:)+B(:).
\item triad: \verb.A(:)=B(:)+s*C(:).
\end{itemize}
on double precision (DP) vectors
\verb.A., \verb.B., and \verb.C. at an array length that is large
compared to all cache sizes. The standard Fortran code allows some
variations as to how the data is allocated. If the arrays are put into
a COMMON block, a configurable offset (``padding'') can be inserted so
that their base addresses vary with the offset in a defined way:
\begin{lstlisting}
PARAMETER (N=20000000,offset=0, &
           ndim=N+offset,ntimes=10)
DOUBLE PRECISION a(ndim),b(ndim),c(ndim)
COMMON a,b,c
\end{lstlisting}
Performance results are reported as bandwidth numbers (GB/s)\@. The
required cache line read for ownership (RFO) on the store stream is
not counted, so the actual data transfer bandwidth for, e.g., STREAM triad
is a factor of 4/3 larger than the reported number (some architectures provide means to
bypass the cache on write misses or claim ownership of a cache line
without a prior read)\@.

\begin{figure}[tbp]
\includegraphics*[width=\columnwidth]{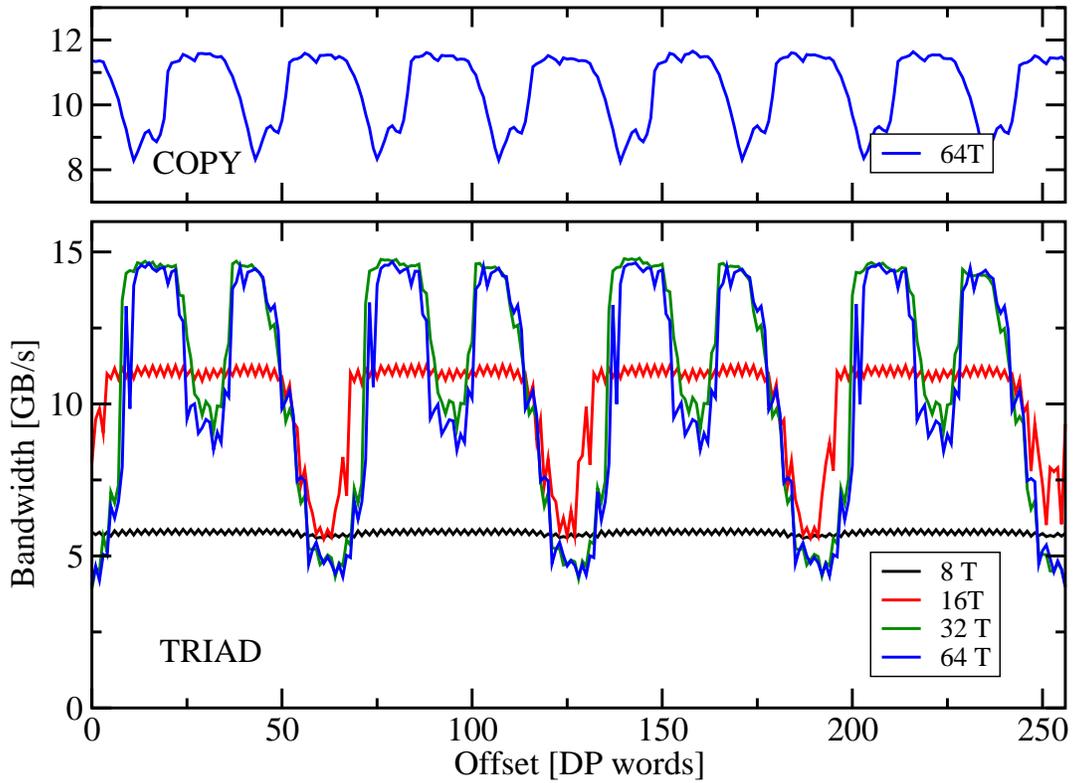}%
\caption{\label{fig:stream1}Lower panel: Parallel STREAM tri\-ad band\-width at 
	$N=2^{25}$ and sta\-tic OpenMP sched\-ul\-ing (no chunk\-size) 
	for thread counts between
	8 and 64 versus array offset on T2. Upper panel: 
	STREAM copy bandwidth for 64 threads.
	Threads were distributed equidistantly
	across cores.}
\end{figure}

Fig.~\ref{fig:stream1} (lower panel) shows STREAM triad performance on 8, 16, 32 
and 64 threads at an array size of $N=2^{25}$\@. 
There is a striking periodicity of 64 for 16 threads and above, and
the 32 and 64 thread data shows an additional, albeit weaker variation
with a period of 32\@. For this simple bandwidth-bound benchmark it
does not seem possible to draw advantage from the T2's large memory
bandwidth. \add[GH]{The reasons for this shortcoming
are as yet unclear; the processor does definitely not suffer from 
a lack of outstanding references as peak bandwidth does not change
when going from 32 to 64 threads. It has been shown, however, that 
kernels which are almost exclusively dominated by loads
can achieve somewhat larger bandwidths \cite{nersc08}, which
leads to the conclusion that at least part of the problem is caused
by overhead for bidirectional transfers. This conjecture is substantiated
by the significantly lower STREAM copy performance 
(upper panel in Fig.~\ref{fig:stream1})\@.}
 
Performance starts off on a very low level at zero offset
and returns to the same level at an offset of 64 which corresponds to
512 bytes. Considering that the array length is a power of two, the
512-byte periodicity reflects perfectly the mapping between memory
addresses and memory controllers which is based on bits 8 and 7\@.
Therefore, the starting addresses of arrays \verb.A., \verb.B., and
\verb.C.  are mapped to the same memory controller if the offset is
zero or a multiple of 64 DP words. This is even true for each single
OpenMP chunk, which means that all threads hit exactly one memory
controller at a time. As the loop count proceeds, successive
controllers are of course used in turn, but not concurrently. At odd
multiples of 32, the situation is improved because bit 8 is different
for array \verb.B.'s base and thus two controllers are addressed, leading to
an expected performance improvement of 100\,\%\@. The fact that 16
threads seem to suffer less under such conditions might be
attributed to congestion effects. 

Finally, at ``skewed'' offsets the addresses of different streams in
one thread and also between threads ensure a rather uniform
utilization of all four memory controllers. Surprisingly, this
condition seems to hold in an optimal way for only about half of all
offsets.  On the other hand one could argue that using an array length
of $2^{25}$ and powers of two for thread counts are choices bound to
provoke aliasing conflicts. In order to show that aliasing must be
taken into account also in the general case, we turn to a more
flexible framework for bandwidth assessment using a self-developed
vector triad code, a 2D relaxation solver and finally an
implementation of the lattice-Boltzmann algorithm.

\SubSection{Vector triad}\label{sec:vt}

\begin{figure}[tbp]
\includegraphics*[width=\columnwidth]{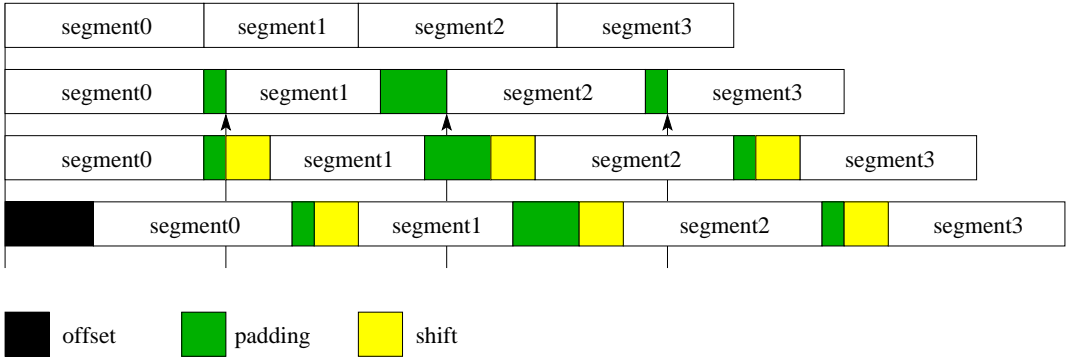}%
\caption{\label{fig:container}Pa\-ram\-eters for align\-ments
	and off\-sets of ar\-ray seg\-ments in the \texttt{seg\_array}
	data structure.}
\end{figure}

The vector triad is quite similar to the STREAM triad benchmark but
features three instead of two read streams
(\verb.A(:)=B(:)+C(:)*D(:).)\cite{schoenauer}\@. We have developed a
flexible C++ framework in which all arrays and also OpenMP chunks can be
aligned on definite address boundaries and then shifted by
configurable amounts (see Fig.~\ref{fig:container})\@.
The array base is aligned to some boundary by allocating memory using
the standard \verb.posix_memalign(). libc function (leftmost border in
Fig.~\ref{fig:container})\@.  The data is then divided into segments,
not necessarily of equal size, and padding is inserted in order to
align each segment except the first to another specific boundary
(arrows in Fig.~\ref{fig:container})\@. After that, a constant amount
of additional padding (``shift'') is added before each segment and
finally the whole data block is shifted by some offset.
Thereby it is, e.g., possible to align an array to a memory page and
then shift a segment that would be assigned to thread $t$ by $t\cdot
128$ bytes.  
\begin{figure}[tbp] 
\includegraphics*[width=\columnwidth]{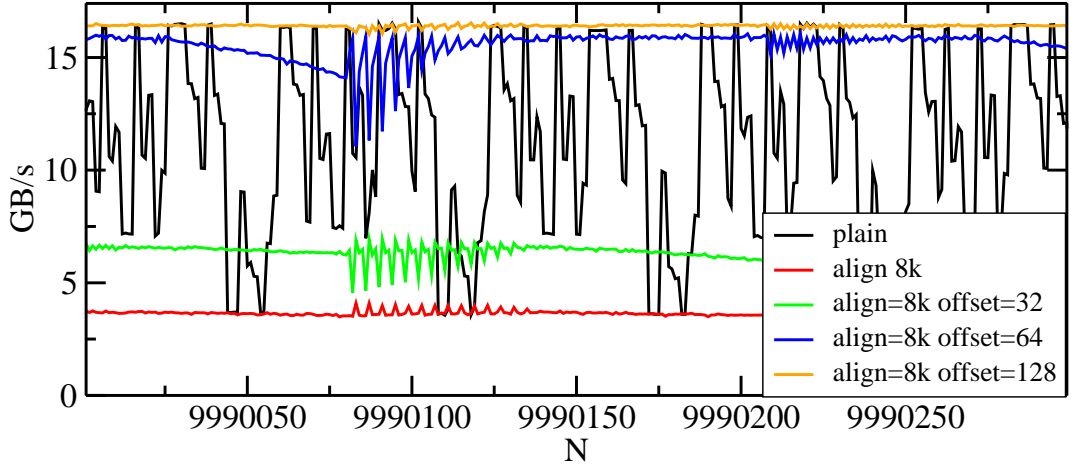}%
\caption{\label{fig:talign}Vector triad performance (64 threads) vs.\ array length
        for plain arrays with no restrictions, 8\,kB alignment for
        all arrays and different byte offsets for array base addresses
	(arrays \texttt B, \texttt C and \texttt D are shifted by
	one, two, and three times the indicated offset, respectively).}
\end{figure}

\add[GH]{Although the segmented data structure can be
equipped with a standard bidirectional iterator, its use is discouraged
in loop kernels because of the required conditional branches in, e.g.,
\texttt{operator++()}\@.} \remove[GH]{The ``gaps'' that appear in the formerly continuous data
stream} \add[GH]{Instead, low-level loops} are handled by a C++ programming technique called
\emph{segmented iterators} \cite{austern01,stengel07} \add[GH]{which 
enables the design of STL-style generic algorithms with performance
characteristics equivalent to standard C or Fortran versions. OpenMP
parallelization directives are applied to the loop over all segments,
and a separate function is called to handle a single segment:}
\begin{figure}[tbp] 
\includegraphics*[width=\columnwidth]{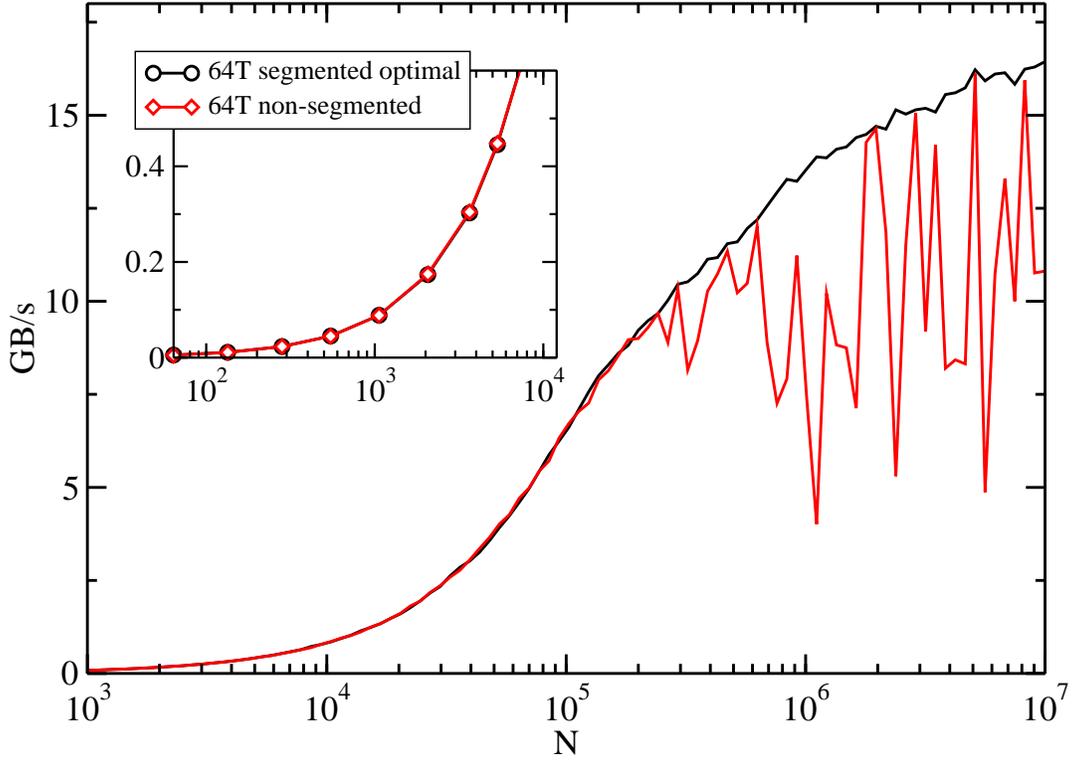}%
\caption{\label{fig:sipenalty}Performance overhead of segmented 
	iterators vs. plain OpenMP (64 threads)\@. For clarity, 
	not all values of $N$ were scanned. Inset: Enlarged 
	small-$N$ region.}
\end{figure}
\begin{lstlisting}[language=C++]
seg_array a,b,c,d; // parameters omitted
...
typedef seg_array::iterator it;
typedef seg_array::local_iterator lit;
typedef seg_array::segment_iterator sit;
sit ai = a.begin().segment();
// ... same for bi, ci, di
#pragma omp parallel for schedule(...)
for(int s=0; s < N_SEGMENTS; s++) {
  lit alb = (ai+s)->begin();
  lit ale = (ai+s)->end();
  lit blb = (bi+s)->begin();
...
  triad(alb, blb, clb, dlb, ale);
}
\end{lstlisting}
\add[GH]{The \texttt{triad()} function performs the actual low-level array
operations and is actually a generic dispatching
algorithm that can handle both segmented and local iterators. Details
about the template mechanism and its general use for high performance
kernels are omitted for brevity and will be published 
elsewhere \cite{segiter08}\@.}

Although C++ is used for administrative purposes, the (purely
serial) inner benchmark kernel can be written in C or Fortran and even
compiled separately without OpenMP, so as to produce the possibly most
efficient machine code.
\change[GH]{We have implemented the triad with this technique, choosing the}{In
our implementation of the segmented triad we choose the}
number of segments equal to the number of OpenMP threads and do
manual scheduling with segment sizes of $\lfloor N/t\rfloor +1$ and
$\lfloor N/t\rfloor$, respectively.  Fig.~\ref{fig:talign} shows
vector triad performance in GB/s versus array length, using different
alignment constraints. \add[GH]{The interval on the $N$ axis was
chosen to clearly show essential features without superimposed
small-$N$ startup effects.} In the ``plain'' case, no special arrangements
were made and arrays were allocated as continuous blocks using
\verb.malloc().\@. This results in very erratic performance behaviour
with a periodicity of 64 DP words, showing ``hard'' upper and lower
limits at roughly 16 and $3.7$\,GB/s, respectively. Aligning all
arrays to page boundaries (8\,kB), one can force an especially bad
situation that corresponds to the zero offset case for the STREAM
triad (bottom line)\@. On the other hand, by choosing suitable offsets
for \verb.B., \verb.C., and
\verb.D. (128, 256 and 384 bytes, respectively, in the optimal case), one
can achieve a nearly perfectly balanced utilization of all four memory
controllers that causes no breakdowns at all (top line)\@. In this case
it is not even required to use padding and shifts (see
Fig.~\ref{fig:container}) for the segments as the large number of
streams (three for reading, one for writing) ensures that even the
single thread features optimal access patterns if the offset is chosen
correctly.\footnote{Although of minor importance here, padding to
16-byte boundaries can greatly improve performance of memory-bound
kernels on x86 architectures due to the possible use of non-temporal
stores that bypass the cache on a write miss.}

\add[GH]{The performance overhead incurred by segmented iterators 
is negligible even for tight loops like the vector triad. 
Fig.~\ref{fig:sipenalty} shows a comparison between plain OpenMP
and segmented triad performance with 64 threads. Optimal alignment
was chosen in the latter case.}

\SubSection{2D relaxation solver}

There are cases, however, where the right choice of offset will not 
suffice. \add[GH]{This happens, e.g., when the number of concurrent
load/store streams is not large enough to address all memory controllers
concurrently with a single thread.} 
As an example and an intermediate step towards more complex applications 
we consider a simple 2D Jacobian heat equation solver using a five-point
stencil \add[GH]{on a quadratic $N\times N$ domain}:
\begin{lstlisting}[language=C++]
#pragma omp parallel for schedule(...)
for(int i=1; i < N-1; i++) {
  for(int j=1; j < N-1; j++)
    dest[i][j] = (source[i-1][j] 
                + source[i+1][j]
                + source[i][j-1] 
                + source[i][j+1])*0.25;
}
\end{lstlisting}
\add[GH]{With four loads, one store and four floating-point operations, 
this kernel has an application balance 
(ratio of bytes loaded or stored vs.\ flops) of 10\,bytes/flop, much 
smaller than the vector triad from the previous section (16\,bytes/flop)\@. 
However, three of the four source operands needed at the current index can be
obtained from cache or registers, given that the amount of cache available per thread
is large enough to accommodate at least two successive rows. If this
condition is fulfilled, the actual data transfer to and from memory 
amounts to only 4\,bytes/flop (6\,bytes/flop with RFO)\@. Comparing with the
achievable STREAM copy bandwidth (Fig.~\ref{fig:stream1}) 
of roughly 18\,GB/s (including RFO) one
should expect a performance of about 3\,GF/s, which corresponds
to 750 million lattice site updates per second (MLUPs/s)\@.}

\add[GH]{Implementing the segmented iterator technique is
straightforward. Each source and destination row is a separate
segment which is subject to the alignment options described
in Sect.~\ref{sec:vt}\@. The parallel OpenMP loop runs over
rows so that scheduling can be done in the standard way. The low-level
kernel is parametrized with iterators pointing to the three current source
rows and the destination row (any template syntax is again omitted):}
\begin{lstlisting}[language=C++]
typedef seg_array::iterator it;
typedef seg_array::local_iterator lit;
typedef seg_array::segment_iterator sit;
sit si = source.begin().segment();
sit di = dest.begin().segment();
#pragma omp parallel for schedule(...)
for(int i=1; i < N-1; i++) {
  lit dl = (di + i)->begin();
  lit sa = (si+i-1)->begin();
  lit sb = (si+i-1)->begin();
  lit sl = (si + i)->begin();
  relax_line(dl, sa, sb, sl, N);
}
\end{lstlisting}
\add[GH]{The \texttt{relax\_line()} function is again purely serial:}
\begin{lstlisting}[language=C++]
void relax_line(lit &dl, lit &sa, 
		lit &sb, lit &sl, int N){
  for(int j=1; j < N-1; j++)
    dl[j] = (sa[j] + sb[j]
           + sl[j-1] + sl[j+1])*0.25;
}
\end{lstlisting}
In a 3D formulation, two additional arguments (rows) to
\verb.relax_line().
would be required. The number of segments equals the number
of rows $N$ and is hence not directly connected to the number of
threads $t$\@. \remove[GH]{OpenMP scheduling is done in the standard way here.
Note that this kernel has only a single read and a single write
stream as source rows that have been read can be reused twice
from cache, assuming the cache is large enough.}

\begin{figure}[tbp] 
\includegraphics*[width=\columnwidth]{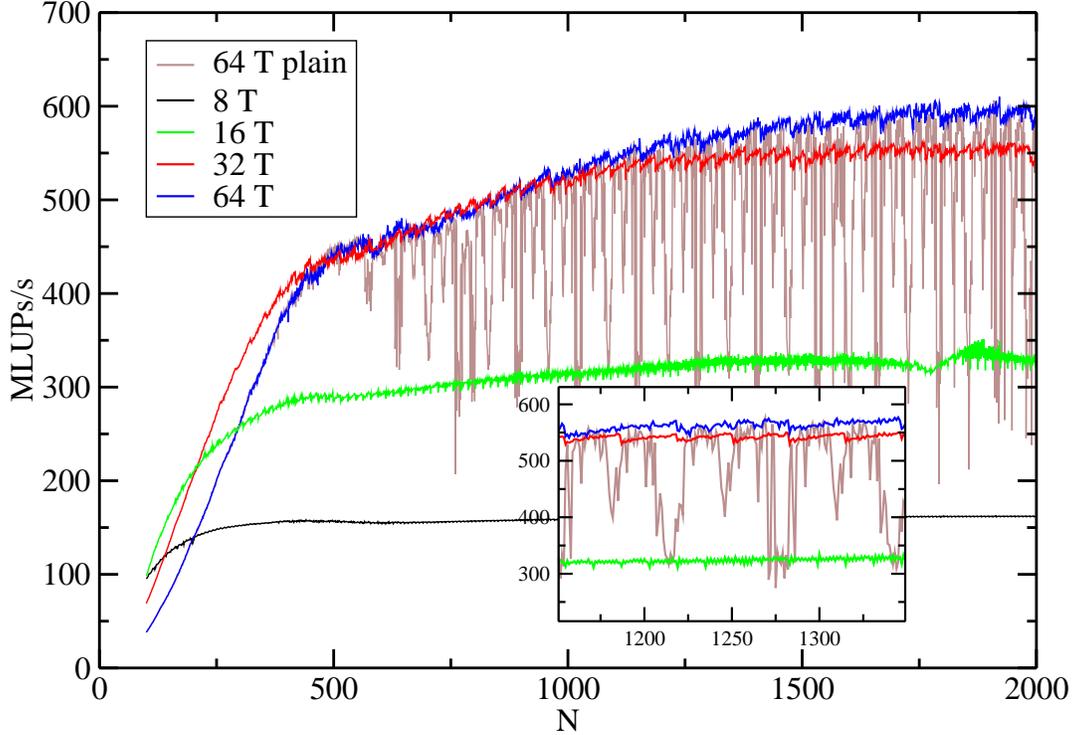}%
\caption{\label{fig:2dgs}Performance and scaling of 2D heat 
	equation relaxation solver versus problem size with optimal 
	alignment and ``static,1'' scheduling. See text for parameters.  
	``Plain'' data with no alignment optimizations is shown for
	reference.}
\end{figure}
Fig.~\ref{fig:2dgs} shows performance results \add[GH]{in MLUPs/s} 
for up to 64 threads using the most optimal set of alignment parameters:
\begin{itemize}
\item Each segment, i.e.\ each row, is aligned to a 512 byte boundary using 
	appropriate padding.
\item By using shift=128, the base addresses of successive segments
	are shifted versus each other so as to address different controllers.
\item As destination row $i$ can only be updated by reading source row 
	$i+1$ first, there is a natural offset between read and write
	streams and no further provisions are required to make sure
	that they address different controllers if shift=128\@.
\item An OpenMP schedule of ``static,1'' has to be used for optimal performance. 
	This is because the 4\,MB L2 cache of the processor is too
	small to accommodate a sufficient number of rows when using 64
	threads if the addresses are too far apart\add[GH]{, i.e.\ if
	the domain is too large}.
\end{itemize} 
\add[GH]{Note that these parameters are the same for all problem sizes 
and can be obtained by analyzing the data access
properties of the loop kernel, together with some knowledge about
the mapping between addresses and memory controllers. No ``trial
and error'' is required. The maximum performance of about 600\,MLUPs/s
is just 20\,\% below expectations from STREAM copy bandwidth.}

For reference, 64-thread data with no optimizations is included. The
typical periodicity of 64 or 32 versus $N$ is clearly visible in the
latter case. The residual ``jitter'' on the optimized data, especially
for large thread counts, is due to the number of rows not being a
multiple of the number of threads. This effect can be expected to
become more pronounced in the 3D case and will be discussed in the
next section on lattice-Boltzmann.

\SubSection{Lattice-Boltzmann algorithm (LBM)}

The advantage of using a special data structure to address alignment
problems is its generality and applicability to non-regular problems
(e.g., segments of different size)\@. It is, however, in some cases
possible to circumvent aliasing effects just by choosing the right
data layout. As an example we consider a lattice-Boltzmann benchmark
that has been developed out of a production code in order to study
various optimizations \cite{wzdh06}\@. For these tests we use a 
3D model with 19 distribution functions (D3Q19) on a cubic domain 
with two disjoint grids (or ``toggle arrays'')\@. \add[GH]{There is a choice as
to which data layout to employ for the cartesian array holding the
distribution functions.} On cache-based architectures the 
propagation-optimized ``IJKv'' data layout\add[GH]{, often 
referred to as ``structure of arrays'',} is 
usually the best choice where I, J and K are cartesian
coordinates and v denotes the distribution function index.
The computational kernel using this
layout is sketched in Listing~\ref{lst:lbm}\@.
\add[GH]{Evidently, the 19 read and 19 write streams are
traversed with unit stride in this case}. 
\begin{lstlisting}[frame=lines,caption={Computational kernel for the IJKv layout  D3Q19 LBM.},label=lst:lbm,float=tbp,belowcaptionskip=\bigskipamount]
real*8 f(0:N+1,0:N+1,0:N+1,0:18,0:1)
logical fluidCell(1:N,1:N,1:N)
real*8 dens, ne, ...
!$OMP PARALLEL DO PRIVATE(...)
do z=1,N
 do y=1,N; do x=1,N
  if ( fluidCell(x,y,z) )  then
   ! read distributions from local cell
   ! and  calculate moments
   dens=f(x,y,z,0,t)+f(x,y,z,1,t)+ &
        f(x,y,z,2,t)+...
   ...
   ! compute non-equilibrium parts
   ne0=...
   ...
   ! write updates to neighbouring cells
   f(x  ,y  ,z  , 0,tN)=f(x,y,z, 0,t)*...
   f(x+1,y+1,z  , 1,tN)=f(x,y,z, 1,t)*...
   ...
   f(x  ,y-1,z-1,18,tN)=f(x,y,z,18,t)*...
  endif
 enddo; enddo
enddo
!$OMP END PARALLEL DO
\end{lstlisting}

Judging from the achievable \change[GH]{triad}{STREAM copy} memory bandwidth
(\change[GH]{over 21}{$\approx$18}\,GB/s including RFO\remove[GH]{, see 
Fig.~\ref{fig:talign}}) and the required
load/store traffic for a single lattice site update (456\,bytes
including RFO), one would expect an LBM performance of roughly
\change[GH]{46}{40}\,MLUPs/s \remove[GH]{(lattice site updates per second)}\@. These kinds of
estimates usually give good approximations for standard multi-core
architectures \cite{diw} \add[GH]{if the kernel is really memory-bound}. 
\remove[GH]{, although the relation between read and
write streams is quite different for both codes (4:1 for triad and
2:1 for LBM)\@.}

\begin{figure}[tbp] 
\includegraphics*[width=\columnwidth]{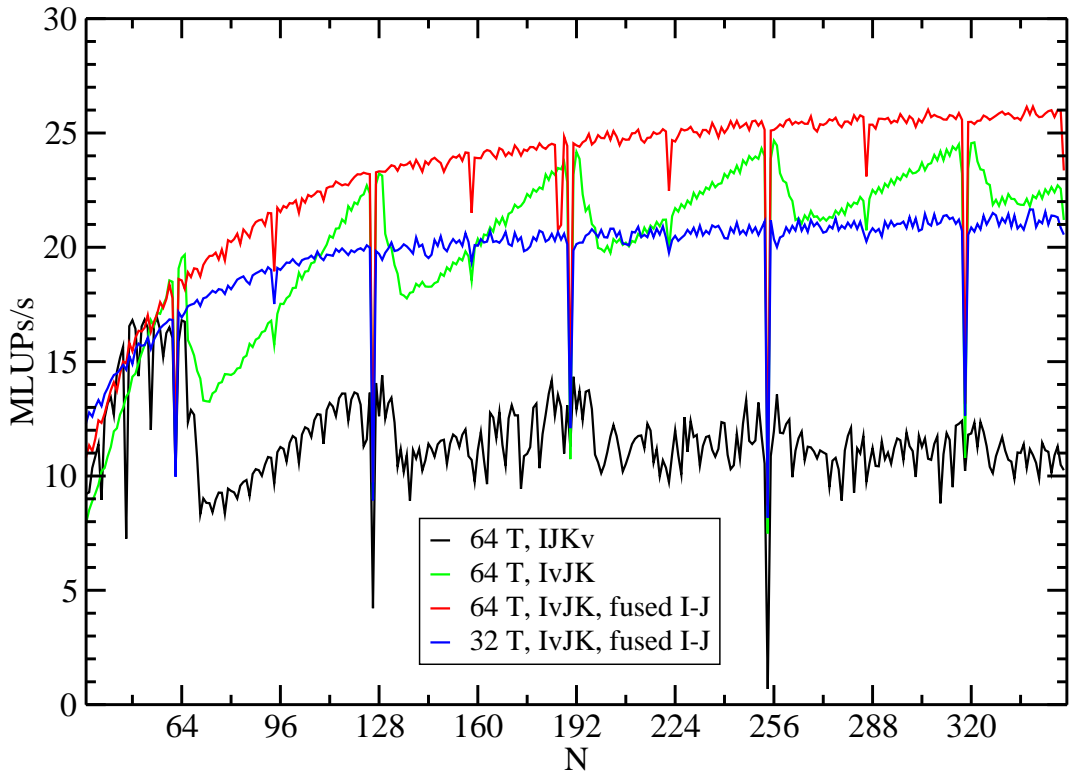}%
\caption{\label{fig:lbmal}LBM performance versus do\-main size (cubic) 
	at up to 64 threads using dif\-fe\-rent da\-ta lay\-outs and scheduling me\-tho\-do\-lo\-gies. 
	The ``modulo variation'' can be eliminated by coalescing the outer
	loop pair (top curve).}
\end{figure}

Fig.~\ref{fig:lbmal} shows performance results \add[GH]{in MLUPs/s}
for LBM on a cubic domain of extent \change[GH]{$N_x\times N_y\times N_z$}{$N^3$} for the
standard IJKv layout as well as for an alternative IvJK
layout. Obviously the latter choice yields twice the performance than
IJKv and also smoother behaviour over a wide range of domain sizes. As
the loop nest is parallelized on the outer level, the fortunate number
of 19 distribution functions leads to an automatic skew between
streams when doing the 19 neighbour updates. The large number of
concurrent stride-1 write streams is of course instrumental in
achieving this effect.

There are two residual peculiarities worth noting. First, if the 1D
domain size is a multiple of 64 (minus two boundary layers), the
well-known cache thrashing effects are ruinous. This could be
eliminated by padding the first array dimension. Second, the
sawtooth-like performance pattern is a ``modulo effect'' which emerges
from \change[GH]{$N_z$}{$N$} not being a multiple of the number of threads. A simple way
to remove the pattern is to coalesce several outer loop levels in
order to lengthen the OpenMP parallel loop. Results for up to 64
threads and two-way coalescing are also shown in Fig.~\ref{fig:lbmal}
and corroborate the call for extensions of the OpenMP standard towards
more flexible options for parallel execution of loop nests.

However, \change[GH]{as Fig.~\ref{fig:lbmal} shows,}{even when
these optimizations are employed,} the
system falls short of \change[GH]{this expectation}{the performance
expectations derived from STREAM} by \remove[GH]{nearly} a factor of 
\change[GH]{two}{1.5}\@. As
for the reason one may speculate that the T2's arithmetic units are a
limiting factor due to the rather low code balance of LBM of $\approx
2.5$\,bytes/flop\add[GH]{, so that the code is not memory-bound on this
processor. This 
conclusion is supported by the observation that LBM performance
does not change if the benchmark is carried out in single precision
(the SPARC core's peak performance is identical for single and
double precision)\@.} 
More cores or a larger peak performance per core
should thus improve the results.

Interestingly, comparing 32- and 64-thread performance in Figs.\
\ref{fig:stream1}, \ref{fig:2dgs} and \ref{fig:lbmal} we conclude that
the smaller the application balance in bytes/flop the larger the gain
when using 64 instead of 32 threads. This is contrary to expectations
as strongly memory-bound kernels should benefit from a larger number
of outstanding references.

\Section{Conclusions}

We have pinpointed aliasing conflicts when accessing memory on Sun's
UltraSPARC T2 multi-core processor. Due to the simple mapping of
memory controllers to physical addresses, bandwidth-intensive code
tends to show large performance fluctuations with respect to problem
size. Using explicit alignment and padding techniques we were able to
remedy aliasing conflicts for a simple vector triad benchmark and a 2D
Jacobi heat equation solver. For a D3Q19 lattice-Boltzmann algorithm 
we could show that an appropriate choice of data layout removes most
of the aliasing. 
\add[GH]{We believe these optimizations to be very relevant on
large-scale systems because predictable one-node performance is essential for
getting good parallel efficiency.}

\change[GH]{In general}{Finally} one must emphasize that in the light of upcoming massively 
multi-core, multi-threaded designs, the rigid OpenMP programming model
might not be the ultimate solution for shared-memory parallel programming
in the future.

\section*{Acknowledgements}

We wish to thank Rick Hetherington, Denis Sheahan and Ram Kunda from Sun 
Microsystems, and Samuel Williams from UCB for valuable discussions.

\end{document}